\documentclass[preprint,12pt,authoryear]{elsarticle}

\usepackage{amssymb}
\usepackage{amsmath}
\usepackage[section]{placeins}

\journal{Astronomy and Computing}

\begin{document}

\begin{frontmatter}



\title{A Multi-modal Fusion Network for Star-Galaxy Classification from CSST Simulated Datasets}


\author[label1,label2]{Zhuoming Han}
\author[label1,label2]{Tianmeng Zhang\cormark[1]}
\cortext[1]{Corresponding authors}
\ead{zhangtm@nao.cas.cn}

\author[label1,label2,label4,label5]{Chao Liu}
\author[label1]{Chenxiaoji Ling}

\affiliation[label1]{organization={National Astronomical Observatories, Chinese Academy of Sciences},
            city={Beijing},
            postcode={100101}, 
            country={People's Republic of China}}
\affiliation[label2]{organization={School of Astronomy and Space Science, University of Chinese Academy of Sciences},
            city={Beijing},
            postcode={100049},
            country={People's Republic of China}}
\affiliation[label4]{organization={Institute for Frontiers in Astronomy and Astrophysics of Beijing Normal University},
            city={Beijing},
            postcode={100875},
            country={People's Republic of China}}
\affiliation[label5]{organization={Zhejiang Lab},
            city={Hangzhou},
            postcode={311121},
            country={People's Republic of China}}
\begin{abstract}
The distinction between stars and galaxies is a fundamental problem in the field of celestial classification. This issue has become challenging for these ongoing and upcoming digital surveys, which will produce terabytes and even petabytes of astronomical data. While deep learning offers a powerful solution for star-galaxy classification in large-scale datasets, most current approaches are limited by their reliance on catalog data alone, which consists primarily of multi-band magnitudes and imprecise morphological parameters. Therefore, we utilize China Space Station Telescope (CSST) simulation data to build a dataset with both image and photometric catalog, including 32,371 stars and 93,525 galaxies. A supervised deep learning network based on ResNet-50 and BiLSTM is proposed to improve the classification of two types of astronomical objects. The features of the catalog and image are integrated by the model, achieving 99.81\% recall for galaxies and 99.66\% recall for stars after training on GPU for 50 epochs. We evaluated the effects of data augmentation and multi-modal data fusion, which demonstrate that our model has commendable performance. Furthermore, our model also has a high accuracy rate for faint astronomical objects and high redshift galaxies, demonstrating its applicability to the upcoming CSST scientific data.
\end{abstract}



\begin{keyword}
Deep learning \sep Multi-modal data fusion \sep Image classification \sep CSST \sep Star \sep Galaxy

\end{keyword}

\end{frontmatter}

\newpage

\section{Introduction}

The categorization of astronomical objects continues to be a significant challenge in observational astronomy, where a key issue is the clear distinction between stars and galaxies. This is because stars and galaxies represent different astrophysical phenomena with important implications for understanding the structure and evolution of the universe. Moreover, the clear distinction between stars and galaxies enhances the accuracy of subsequent astrophysical research, both in theory and observation.

Traditional classification methods are highly dependent on the professional expertise and experience of astronomers. However, with the advent of numerous ongoing and upcoming astronomical surveys, such as the Dark Energy Survey (DES; \citealt{1}), the Sloan Digital Sky Survey (SDSS; \citealt{2}), the Roman Space Telescope \citep{3}, Euclid \citep{4}, the Zwicky Transient Facility (ZTF; \citealt{5}), the Panoramic Survey Telescope and Rapid Response System (Pan-STARSS; \citealt{6}), The Legacy Survey of Space and Time (LSST; \citealt{7}), Gaia survey \citep{8} and China Space Station Survey Telescope (CSST; \citealt{9}) etc., the situation has evolved into a more complex state. These surveys are expected to obtain imaging data on millions to billions of stars and galaxies, making manual analysis and classification impractical. Specifically, the LSST will produce about 15 TB of raw data per night and 32 trillion observations of 40 billion objects \citep{7}, and the CSST will map approximately 17,500 square degrees of the sky, obtaining photometry for more than a billion galaxies and a billion stars \citep{21-2}. Therefore, it is necessary to develop methods to accurately and rapidly distinguish between stars and galaxies.

Currently, three primary methods are commonly used for classifying astronomical objects. The first method is to analyze the spectral differences \citep{9-1, 9-2}. Stellar spectra typically exhibit distinct absorption lines, caused by various elements in the stellar atmosphere that absorb specific wavelengths of light. In contrast, the spectra of galaxies may display both absorption and emission rays, especially if there is an active galactic nucleus (AGN) or a large amount of ionized gas present in the galaxy. The second method is based on morphological differences. Generally, stars are described by the point spread function (PSF), while galaxies exhibit extended, diffuse structures \citep{10, 11, 12}. This method is consistent with the classification of spectral differences \citep{13} and widely used within SExtractor (Source Extractor; \citealt{2011ASPC..442..435B}). The third method is to classify stars and galaxies according to their different positions on the color-color diagram \citep{14, 15, 16}. An approach that combines these methods should be more effective in maximizing the utilization of available data \citep{17, 18, 19}.

Although identifying the types of astronomical sources may not be inherently difficult through spectroscopy, the process becomes complex and time-consuming when attempting to gather such detailed observations for millions of individual sources \citep{19-1}. Classifying sources using images and photometric catalogs across multi-bands, and assigning labels based on their color indices and morphological structure, is a significantly more expeditious approach. \citealt{22} used 5-band images and the carefully selected photometric parameters from SDSS Data Release 16, obtaining an overall accuracy of 98.1\% for star-galaxy classification, which performs better than the individual model for both classifications. Several recent studies also employed the modal fusion method and achieved notable classification outcomes \citep{19-4, 22-3}.

In recent years, the classification of stars and galaxies in astronomy has been approached using an assortment of Machine Learning (ML) algorithms \citep{19-2, 19-3, 19-4}. They can identify complex nonlinear behavior in the multi-dimensional feature space. Among them, Deep Learning (DL) networks perform better on some intricate issues with more parameters and more complex structures, such as Convolution Layers, Pooling Layers, and Fully Connected Layers, which can exploit the potential information in celestial images. 

\citealt{20} proposed a method to distinguish between stars and galaxies using high-resolution images from the Hubble Space Telescope (HST). However, this method utilized only the $i$-band image data and did not incorporate any multi-band features. When the flux is relatively weak, the galaxies tend to be more compact, making accurate categorization challenging, even using their polar transformation method \citep{20-1}. \citealt{21} applied the random forest method to the classification of stars and galaxies in the S-PLUS 12 band catalog and achieved 95\% accuracy. The importance of various features in the catalog for classification results was explored, and it was concluded that morphological parameters are relatively more important, especially the full width at half maximum (FWHM) and PSF, which could be extracted from the images as morphological features in these catalogs.

Most existing methods rely on single-band catalogs or images, which inherently contain less information than their multi-band counterparts. Multi-band image data provides much more information about astronomical objects, revealing their morphology in each band and facilitating the distinction of stars from galaxies. 

In this paper, we aim to classify the objects detected from the main survey simulation data of the CSST, covering 0.75 $deg^2$ with seven bands. The DL algorithms that we employ in this work are supervised, with both multi-band images and photometric catalogs well integrated to better distinguish between stars and galaxies.

This paper is structured as follows: we introduce the dataset used in our study in Section 2. In Section 3, we describe the pre-processing work applied to these data. In Section 4, we first propose the framework of the method and introduce its implementation details. In Section 5, we present the classification results of the network. Some ablation experiments are conducted, and the effectiveness of the model in the cases of faint astronomical objects and high-redshift galaxies is verified. The conclusion is given in Section 6.

\section{Data}
\label{sec2}
CSST is a 2-meter aperture space telescope equipped with five instruments, which are multi-band Imaging and Slitless Spectroscopy Survey Camera (MSC), THz Spectrometer (TS), Multi-Channel Imager (MCI), Integral Field Spectrograph (IFS), and Cool Planet Imaging Coronagraph (CPI-C). The MSC is designed to perform multi-band imaging and slitless spectroscopic surveys. The focal plane with 1.1 square degrees of Field of View (FoV) is divided into 30 9k$\times$9k CCD detectors, including seven photometric filters (\textit{NUV, u, g, r, i, z, y}) and three spectroscopic gratings (\textit{GU, GV, GI}). The FoV covered by each detector is about $11 \times 11 ~ arcmin^2$, with a pixel scale of 0.074 arcseconds. The newest released dataset covered $5 \times 5~\deg^2$ has been simulated by the CSST simulation framework, which was built on the GalSim package and incorporates detailed modeling of astronomical sources, instrumental effects, and observational conditions \citep{21-2}. These data were processed and facilitated by the pipeline specifically developed for the CSST survey.

\begin{figure}[t]
\centering
\includegraphics[width=1.0\linewidth]{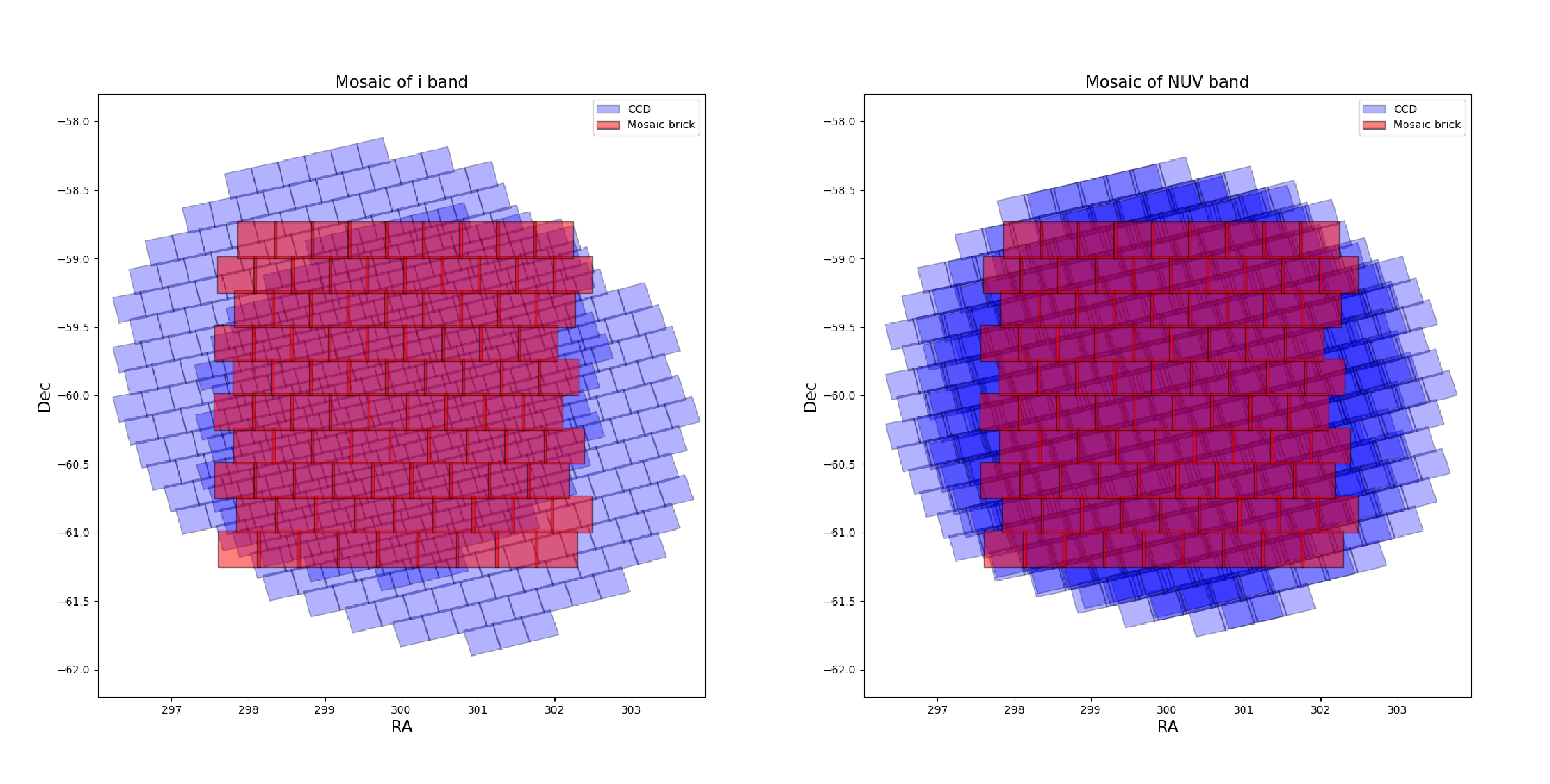}
\caption{The sky region coverage of simulation data. The blue squares represent single exposure images in the $i$-band (left panel) and $NUV$-band(right panel), while the red squares represent mosaic bricks.}
\label{fig-mosaic_NUV+i}
\end{figure}

\subsection{Multi-band Image}
The dataset utilized in this work is one of 5 $\deg^2$ sky regions centered on Right Ascension (RA) = 300.0 deg, Declination (Dec) = -60.0 deg, which contains 7-band images and photometric catalogs. The coverage of these data is presented in Fig. \ref{fig-mosaic_NUV+i}. The $NUV$ and $y$ bands have four detectors, while the $u$, $g$, $r$, $i$, $z$ bands have two detectors each. Each detector forms a brick that covers a certain area of the sky. There is overlap between these detectors, and the images of the overlapping areas are resampled to form a mosaic. The single exposure images with the same filter have been combined to bricks of $16 \times 16$ arcminutes, which are convenient to align with different bands and cutouts.

The multi-band image data in the 7 photometric filters, $NUV$, $u$, $g$, $r$, $i$, $z$, $y$ are extracted from the FITS (Flexible Image Transport System) files obtained in each filter from the CSST Simulation Data, which consists of two to four exposures of the same wavelength that are combined into a Mosaic image. Each mosaic image includes the SCI matrix, the weight matrix, and the flag matrix. The SCI matrix is the 2D data array which contained the normalized flux of astronomical objects in unit e/s. Both SCI and weight matrices are interpolated according to the position of cosmic ray, and the mask of bad pixels is still retained in flag matrix.

\subsection{Multi-band Photometry Catalog}
We use both a multi-band simulation input catalog and photometric catalogs obtained by the CSST data processing pipeline. The input catalog is simulated according to the distribution of various types of astronomical objects based on cosmological theories, which include RA., Dec., object type, and 7 band magnitudes, etc. Almost all spectral types of stars are included in the catalog, while some special stellar types, such as chemically enhanced stars and binary systems, are not included. Galaxy morphology includes elliptical galaxies, spiral galaxies, and pure disk galaxies \citep{21-3}. The photometric catalog contains more detailed parameter information for each source, such as RA., Dec., magnitudes with different apertures, spread model, and isophotal radius, etc., which are generated by CSST data reduction pipeline. Specifically, we obtain the true label from the input catalog and the magnitude of 7 bands from the photometric catalog. The morphology parameters are not taken into account to avoid introducing uncertainties and interference.

\section{Data Preprocessing}
In this section, we provide a brief overview of the preprocessing steps, including cross-matching, image cutouts, and data augmentation, designed to ensure that the data better meet the input requirements of our model. 

\subsection{Cross-match of Catalogs}
The final photometric catalog is created by cross-matching the input catalog and photometric catalog with a maximum positional error of 0.3 arcsec. Each sky region has an input catalog and seven corresponding photometric catalogs in the 7 bands. We used the "or" relationship to match the sources between each pair of catalogs one by one, and then retained the sources that had magnitudes from four or more bands. The magnitudes were max-normalized to the [0, 1] range to ensure stable model training.
 
\subsection{Image Cutouts}
To recognize the image of each source, we need to cut the mosaic image containing tens of thousands of sources into cutout images that contain only one source. Each band mosaic image was scanned to ensure the object was within the pixel boundaries. Finally, we cut our samples into cutouts with the same size of $50 \times 50$ pixels, centering the objects within the image. With a separate cropped image in each band, they were stacked to produce a 7-channel FITS format image.	
	
\begin{figure}[t]
\centering
\includegraphics[width=1.0\linewidth]{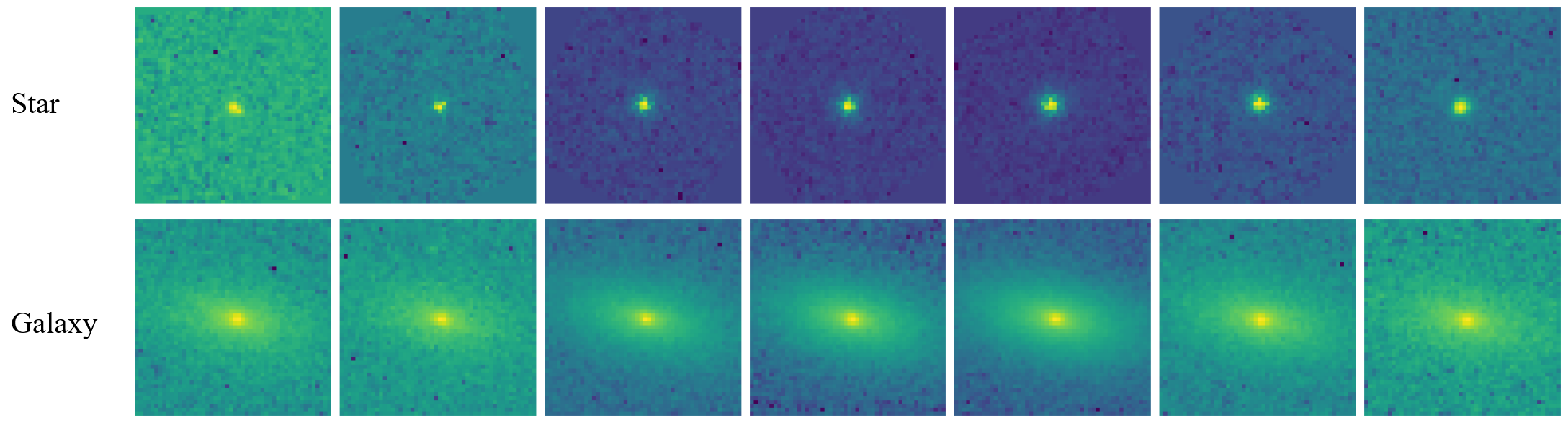}
\caption{Cutout image for 7 bands (from left to right: $NUV$, $u$, $g$, $r$, $i$, $z$, $y$), top: $50 \times 50$ pixels cutout images of star, which appears as point sources convoked with the PSF. bottom: $50 \times 50$ pixels cutout images of galaxy, which appears as extended sources.}
\label{fig00-1}
\end{figure}
	
\subsection{Data Augmentation}

\begin{figure}[t]
\centering
\includegraphics[width=0.7\linewidth]{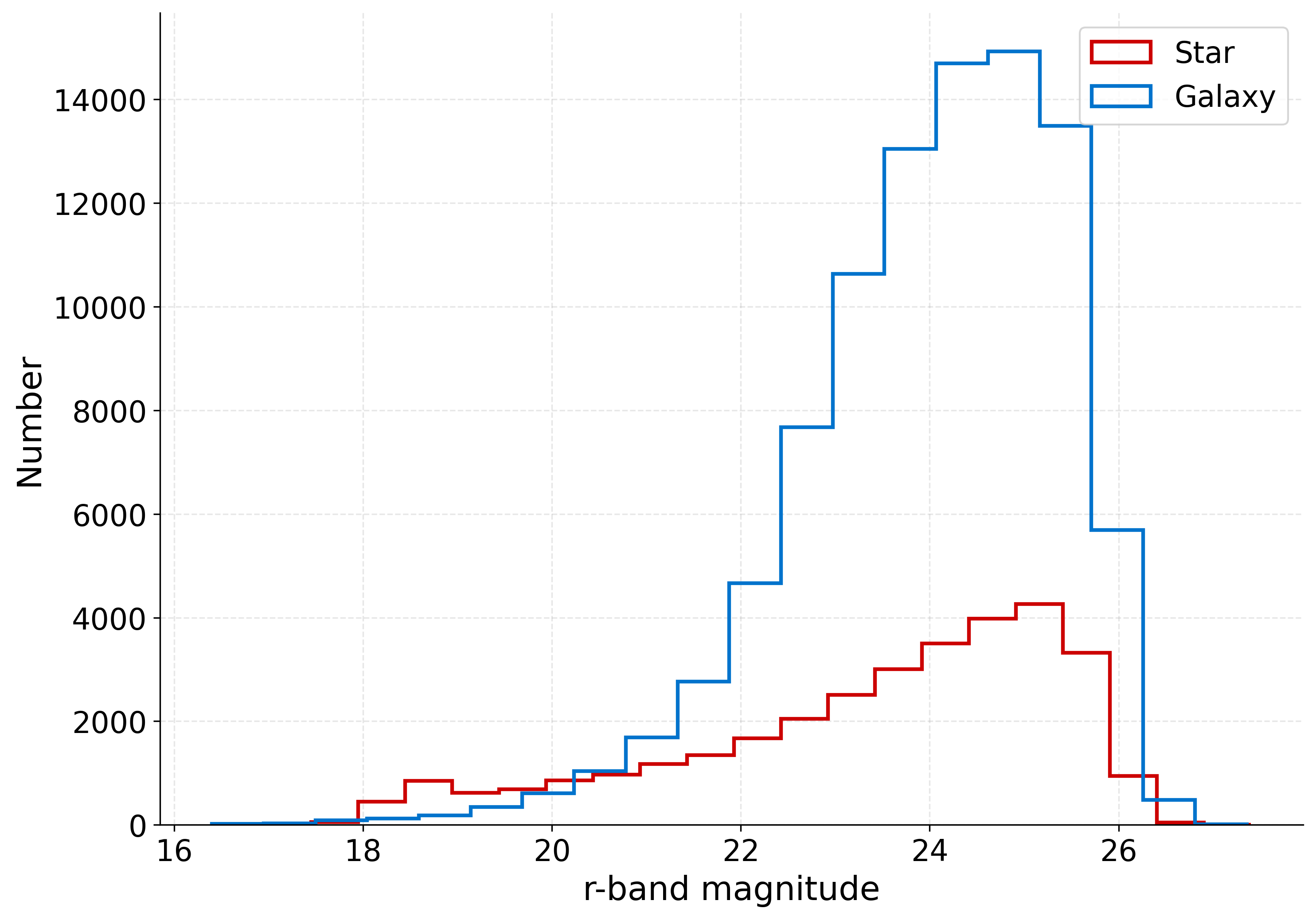}
\caption{The r-band magnitude distribution of stars (red) and galaxies (blue) from photometric catalogs.}
\label{fig00-distribution-r-band}
\end{figure}

Our sample contains 93,525 galaxies within 0.75 $deg^2$, which is three times the number of stars that have 32,371. In Fig. \ref{fig00-distribution-r-band}, we present the distribution of r-band magnitudes for stars and galaxies (the redshift distribution of galaxies is shown in Fig. \ref{fig00-7-2}). Because of the unbalanced celestial samples, it is easy for the classification network to identify more sources as galaxies to achieve higher accuracy. At the same time, it rarely learns the substantive features of the samples. We intend to increase the number of stars to alleviate the current imbalance in the samples. 

Data augmentation is one of the effective methods to increase the number and diversity of samples. There are a variety of standard data augmentation strategies, but not all augmentation techniques are applicable to our specific dataset \citep{21-4}. This is particularly true for methods involving photometry, which are contingent on the quality and characteristics of the original images. This study has adopted three augmentation methods that are better suited to our data. These selected methods have proven to be more effective for enhancing our dataset without compromising the integrity of the photometry results. (1) Random Horizontal Flip. (2) Random Vertical Flip. (3) Rotates randomly between -45 degrees and 45 degrees. The size of the image after rotation is consistent with that of the original image. Pixels beyond the original image area are intercepted and discarded, and 0 padding is used for areas without pixels.
	
We enhanced the star samples by 3 times to make the sample numbers of stars and galaxies almost consistent. To maintain the one-to-one correspondence between the catalog and the image, the original catalog was accordingly copied after the image was enhanced, ensuring that it continued to match the new image data.

\section{Method}
	
\subsection{Data Fusion}

Data fusion effectively combines data from multiple sources to achieve more accurate, complete, and reliable estimates and judgments than a single information source. The core goal is to extract valuable information from multiple sources of data and reduce redundancy, improving system reliability and decision-making ability. Data fusion includes pixel-level fusion, feature-level fusion, and decision-level fusion \citep{21-5}. Different levels of data fusion come with their own advantages and disadvantages, and the choice among them depends on the specific task to be addressed.

The framework 'MargNet' was designed to classify stars, quasars, and compact galaxies using the $u, g, r, i, z$-band images and the corresponding catalog from SDSS Data Release 16 \citep{22}. This framework integrates a convolutional neural network (CNN) with an artificial neural network (ANN) and demonstrates superior performance compared to previous methods. However, catalog and image data are fused at the decision-level, meaning only the final results of each data source are used, and the details and potential information in the data are ignored. In the complex system of DL networks, it is difficult to trace the specific details behind the decision, and the design of rules directly affects the fusion effect.

Some researchers also use the Vision Transformer (ViT) to process the features of celestial images \citep{19-4}. Nevertheless, this method is less capable of extracting the local correlation between pixels compared to CNN-based networks. The self-attention mechanism in the ViT has a computational complexity that is quadratic in terms of the number of input image blocks, resulting in significant computational and memory overhead.

In this paper, a CNN-based network is adopted to extract features from images. Finally, a one-dimensional feature vector is obtained through the convolution flatten layer and fused with the spectral energy distribution (SED) features to obtain a feature vector with richer information. 
	
\subsection{Network Architectur}

\begin{figure}[t]
	\centering
	\includegraphics[width=1.0\linewidth]{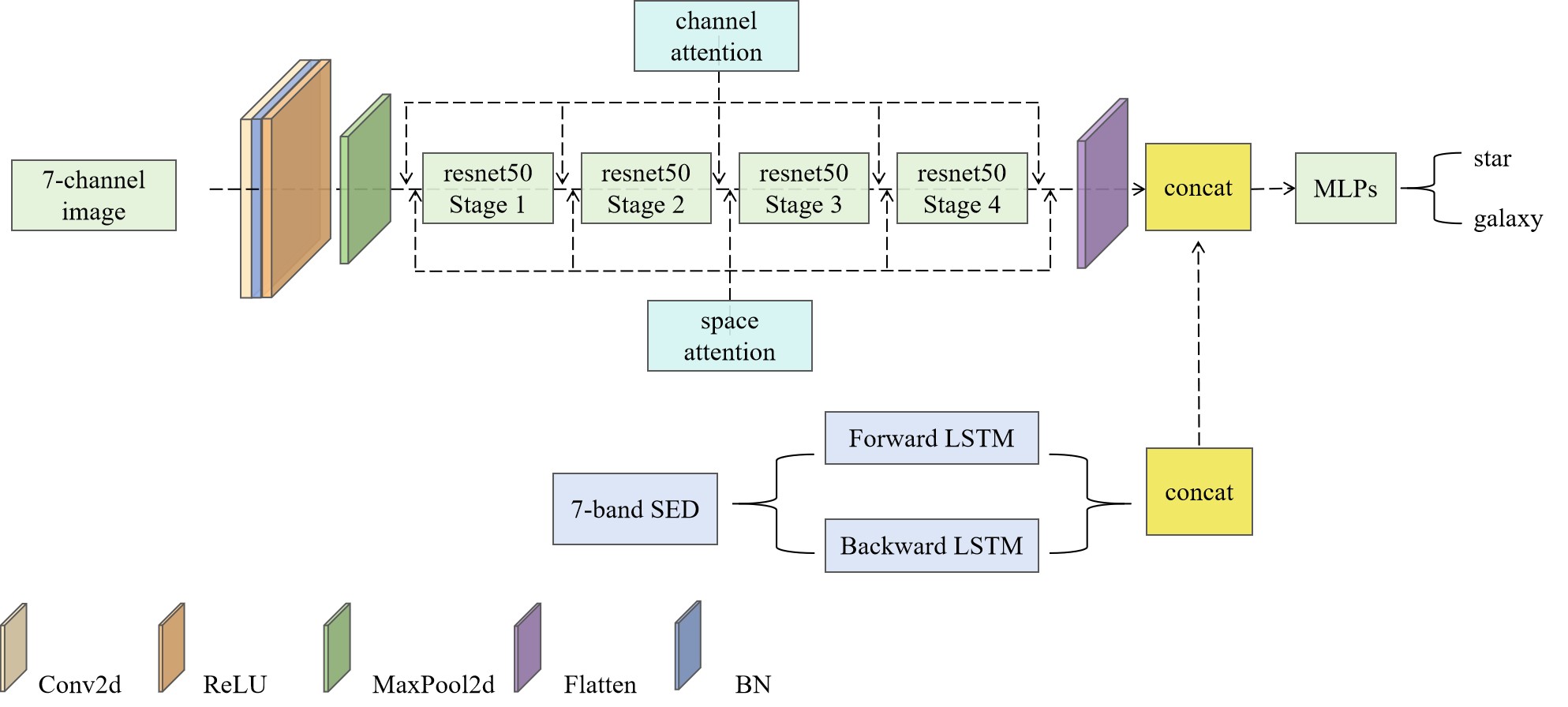}
	\caption{RBiM network. The 7-channel image data and catalog are processed through the feature extraction network, and then concatenated in the vector space and connected to the MLP layer to predict the probabilities of star and galaxy. The main layers used are illustrated in the legend.}\label{fig00-2}
\end{figure}

\begin{figure}[t]
	\centering
	\includegraphics[width=0.9\linewidth]{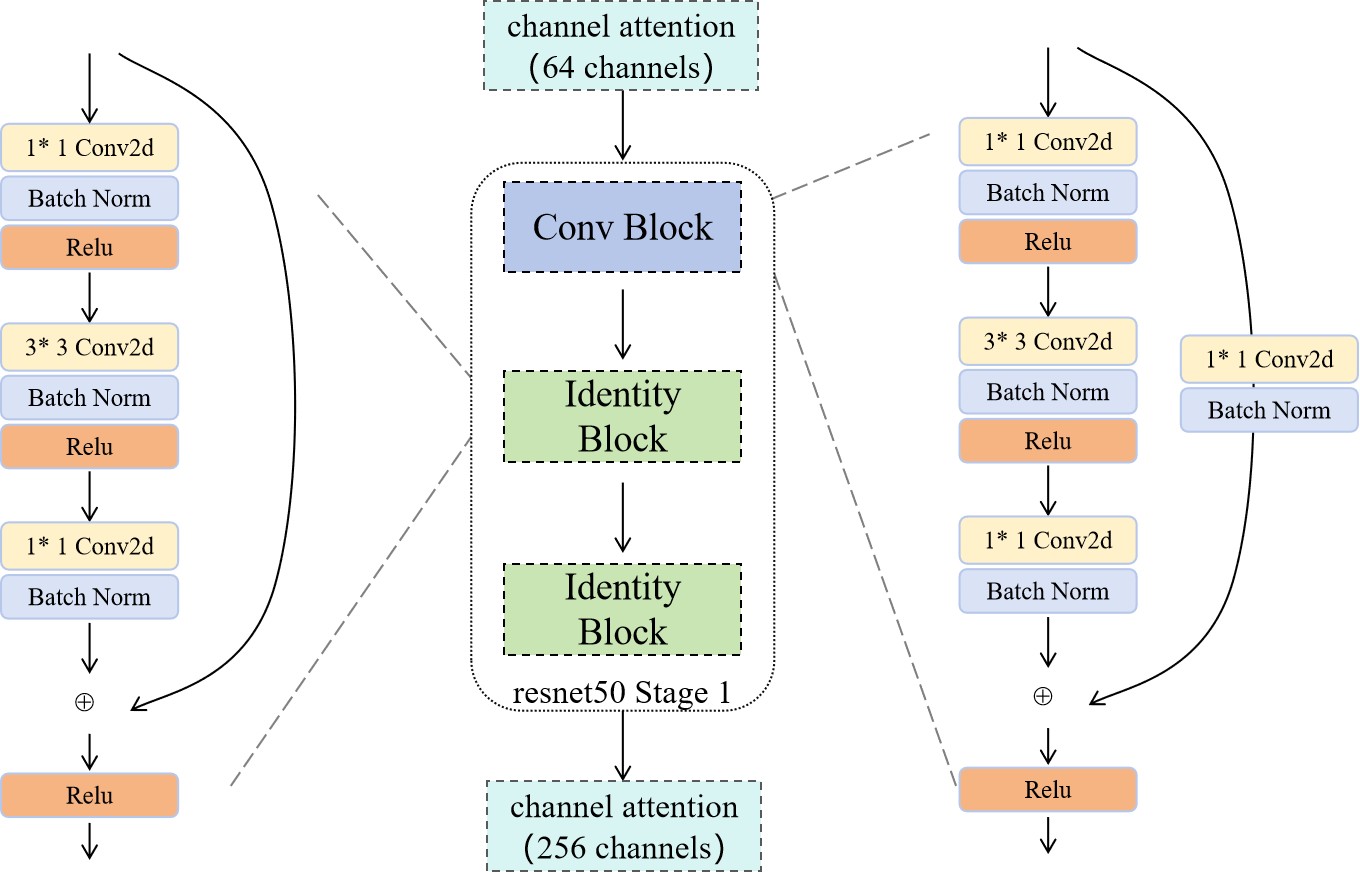}
	\caption{Residual connection in Stage 1 of ResNet-50, including two types of residual blocks, Conv Block and Identity Block. The dimensions of the input and output of Conv Block are different, which can change the dimensions of the network. Identity Block has the same input and output dimensions, which can deepen the network. The attention mechanism is incorporated after each stage to extract more significant image features.}\label{fig00-2-2}
\end{figure}

We developed the RBiM network (a multimodal network combining Res-Net-50 and Bi-directional Long Short-Term Memory (BiLSTM)) using the DL framework, which runs on top of PyTorch \citep{23}. A multi-band image and catalog data fusion network is proposed to achieve better classification of stars and galaxies. The specific structure is shown in Fig. \ref{fig00-2}. ResNet-50, which consists of five stages, is employed as the backbone network \citep{23-2}. The first stage includes an initial convolutional layer, a batchnorm layer, a ReLU activation layer, and a maxpooling layer, which are responsible for extracting features, channel normalization, activation, and downsampling, respectively. The following four stages consist of several bottleneck blocks, each of which is composed of three convolutional layers, such as the structure in Fig. \ref{fig00-2-2}. It employs shortcut connections to facilitate residual learning, which mitigates the training difficulty commonly encountered in traditional deep neural networks as their depth increases.
    
We added an attention mechanism after the first CNN layer and each stage of ResNet-50 in a progressive form, which means that attention modules of different scales are added after each of the four stages of ResNet-50, to obtain the most important feature in the different scale feature maps. The attention mechanism consists of two parts: the channel attention and the spatial attention. Channel attention is used to capture the dependencies among the images of the 7 channels and 1 $\times$ 1, 3 $\times$ 3, and 5 $\times$ 5 convolution kernels of different sizes are used to extract features. Spatial attention uses adaptive weights to capture the most important areas in the feature map.

The catalog feature is obtained by the BiLSTM network, which can extract bidirectional causal information and obtain the correlation between the magnitudes of each band of the samples \citep{23-3}. The BiLSTM consists of two independent Long Short-Term Memory (LSTM) layers. The forward LSTM captures the dependency from the $NUV$ band to the $y$ band, while the backward LSTM captures the dependency from the $y$ band to the $NUV$ band. Then, the outputs of the two are concatenated. This structure effectively captures the distribution of the SED of astronomical objects and has stronger modeling capabilities for sequences containing missing data. Compared to the traditional Multilayer Perceptron (MLP) network, it is better suited to our missing band samples.

As shown in Fig. \ref{fig00-2}, the morphological features of the 7-band cutout images ($7 \times 50 \times 50$) are extracted by the ResNet-50 network with the attention mechanism. Then the fusion of features is carried out using a flat layer and the SED features ($7 \times 1$) to flatten into a vector of dimensions $1 \times 256$. Following the fusion layer, the classification results are generated through three layers of MLP.

\section{RESULTS AND DISCUSSION}
	
\subsection{Evaluation Criteria}
The performance of the trained network was evaluated using the test sample. To provide a comprehensive evaluation of the model, we selected multiple factors, including precision, recall, accuracy, and F1-score. Relying on a single evaluation metric often leads to one-sided or even incorrect conclusions. By examining the model from multiple perspectives using these indicators, we can gain a more thorough understanding of its performance and identify any existing issues.
	
\begin{equation}
Precision = \dfrac{TP}{TP+FP}
\end{equation}

Precision refers to the proportion of the number of correctly classified positive samples (True Positive (TP)) to the number of positive samples (True Positive and False Positive (FP)) by the classifier. 
	
\begin{equation}
Recall = \dfrac{TP}{TP+FN}
\end{equation}
	
Recall refers to the proportion of the number of correctly classified positive samples (True Positive) to the number of true positive samples (True Positive and False Negative (FN)). 
	
\begin{equation}
Accuracy = \dfrac{TP+TN}{TP+TN+FP+FN}
\end{equation}

\begin{equation}
F1-score = \frac{2 \times precision \times recall}{precision + recall}
\end{equation}

Accuracy is the ratio of correctly classified samples to the total number of samples. F1-score is the weighted average of precision and recall. 
	
\subsection{Overall Performance} 
	
We shuffled the dataset consisting of 93,525 galaxies and 32,371 stars after data preprocessing, and then divided it into a training set and a test set in a 9:1 ratio. The test set includes 9,358 galaxies and 3,232 stars. Subsequently, the 3 times of data augmentation is performed on the star samples in the training set. To avoid overfitting, we adopted 3-fold cross-validation, dividing the training set into 3 equal parts, and alternately used 2 of them as training data and the remaining one as validation data for the experiment. In addition, the method of the cross-entropy loss function with label smoothing and gradient clipping was employed to prevent overfitting. Besides, the cosine preheating technique was adopted. In the early stage of model training, a lower learning rate was used. Subsequently, as the number of training epochs increased, the learning rate gradually increased linearly. After that, it slowly decreased in the form of a cosine function. This method prevents rapid parameter updates on small batch samples and is beneficial for model convergence.
    
\begin{figure}[t]
	\centering
	\includegraphics[width=0.7\linewidth]{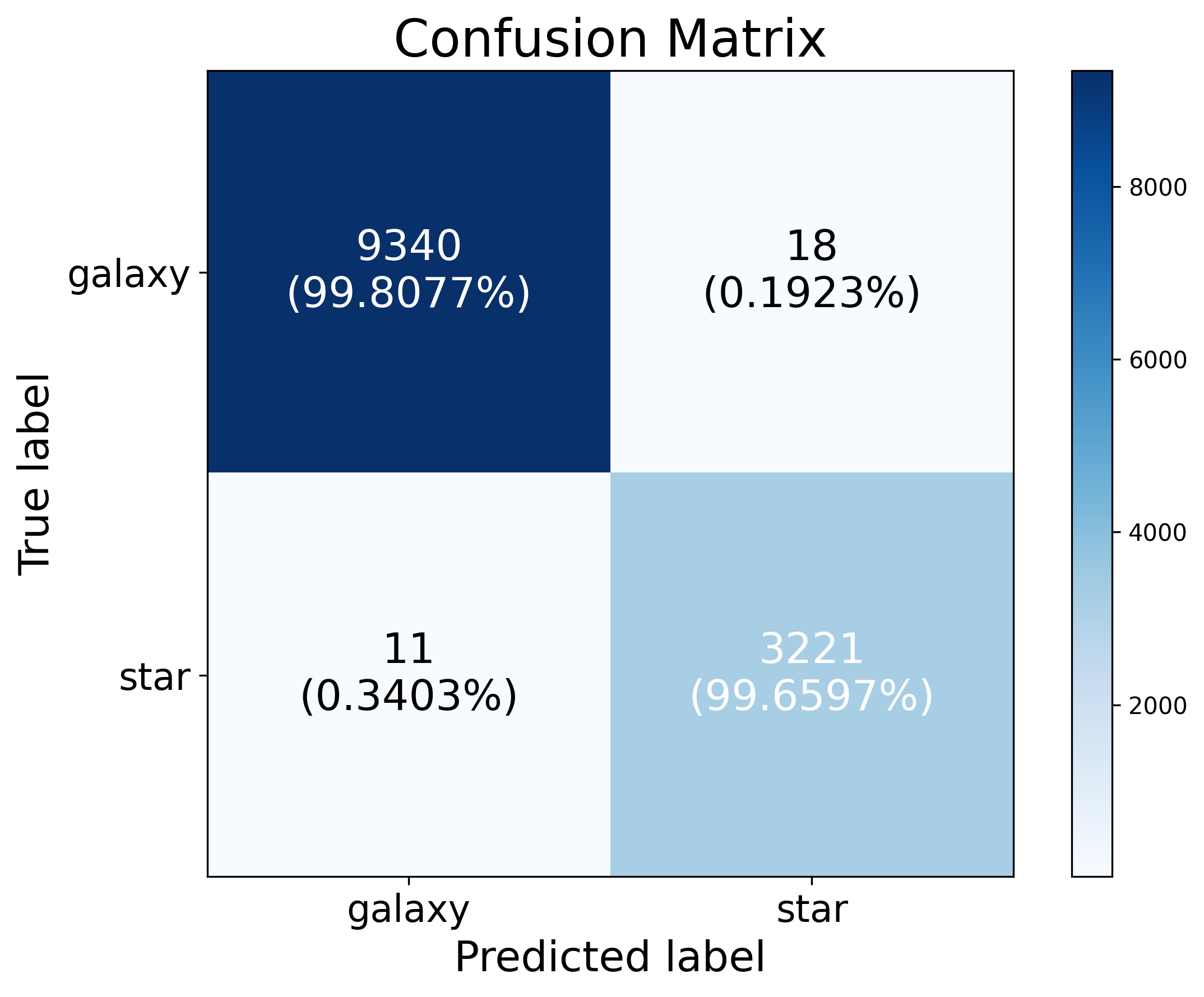}
	\caption{Confusion matrix for classification of stars and galaxies. The vertical axis represents the true labels of the samples, while the horizontal axis shows the classification results of the model. And the darkness of the color indicates the quantity of the samples.}\label{fig00-6}
\end{figure}

Using these strategies and training for 50 epochs, the galaxies achieved a recall rate of 99.81\% and a precision rate of 99.88\%, while the stars achieved 99.66\%, 99.44\%, respectively. The general classification accuracy was above 99.75\%, according to the confusion matrix in Fig. \ref{fig00-6}. Among the 12,590 sources in the test set, only 18 galaxies were wrongly classified as stars, and 11 stars were wrongly classified as galaxies. Some figures illustrating the performance of the model are shown in the appendix, including ROC curves and t-SNE visualization of features. Analysis of the 7-channel mosaic images revealed the presence of interference sources in proximity to certain samples, along with observable star proper motion and mosaic alignment errors in several cases. These rare phenomena may cause stars to stretch out as extended sources, which led to the occurrence of misclassifications.

To test the effectiveness of data enhancement, we performed a comparison experiment with raw data without enhancement. Without data augmentation, the classification recall is 99.76\% for galaxies and 99.50\% for stars. The recall rate for galaxies and stars decreased by 0.05\% and 0.16\%, respectively. This is because the characteristics of the star obtained through the network have been significantly reduced as a result of the extremely small sample size compared to the entire galaxy samples. Moreover, in this situation, classifying some samples that are difficult to categorize as galaxies is beneficial to increasing the overall classification accuracy.

\subsection{Advantages of Multi-modal}

We have quantitatively assessed the performance improvement from multi-modal data fusion by comparing the classification results using datasets with either image or catalog. To obtain more reliable comparison results, we added two sets of experiments, which divide the training set and test set in an 8:2 and 7:3 ratio, respectively. Table \ref{table-1} provides the evaluation criteria values for the three sets of experiments, along with their average values. For galaxies, the recall rate improved by 11.17\% compared to the catalog-only model and by 0.01\% compared to the image-only model. Meanwhile, the precision rate increased by 1.58\% compared to the catalog-only model, and by 0.05\% compared to the image-only model. For stars, the recall rate improved by 4.08\% compared to the catalog-only model and by 0.12\% compared to the image-only model, while the precision rate increased by 25.01\% compared to the catalog-only model and by 0.04\% compared to the image-only model. These results clearly demonstrate that multi-modal data fusion enhances the classification capability of models, especially when using only catalogs. Differences in the performance of various modal methods can also be directly observed from the F1 score.

In addition to the 3-fold cross-validation, we added 5-fold cross-validation and calculated the standard deviation of the accuracy of each fold model on the validation set to better quantify and compare the performance of the models. Both the multimodal and image-only models exhibit a mean standard deviation of approximately 0.01\%, which indicates the stability and reliability of our models. Although fluctuation in accuracy rates can further reduce the performance gap between the two models, the multimodal model still retains the comparative advantage.

\begin{table}[h!]
\centering
\begin{tabular}{c c c | c | c | c}
		\hline
		ratio & type & metrics & catalog-only & image-only & fusion \\
        \hline
        & & recall & 89.96\% & 99.75\% & 99.81\% \\
		& galaxy & precision & 98.53\% & 99.85\% & 99.88\% \\
		9:1 & & f1-score & 94.05\% & 99.80\% & 99.84\% \\
        \cline{3-6}
		& & recall & 96.10\% & 99.57\% & 99.66\% \\
		& star & precision & 76.77\% & 99.29\% & 99.44\% \\
		& & f1-score & 85.35\% & 99.43\% & 99.55\% \\
		\hline
		& & recall & 88.61\% & 99.74\% & 99.75\% \\
		& galaxy & precision & 98.27\% & 99.82\% & 99.89\% \\
		8:2 & & f1-score & 93.19\% & 99.78\% & 99.82\% \\
        \cline{3-6}
		& & recall & 95.49\% &  99.49\% & 99.68\% \\
		& star & precision & 74.35\% & 99.26\% & 99.28\% \\
		& & f1-score & 83.60\% & 99.37\% & 99.48\% \\
        \hline
        & & recall & 87.22\% & 99.78\% & 99.76\% \\
		& galaxy & precision & 98.08\% & 99.80\% & 99.84\% \\
		7:3 & & f1-score & 92.33\% & 99.79\% & 99.80\% \\
        \cline{3-6}
		& & recall & 95.02\% &  99.43\% & 99.53\% \\
		& star & precision & 71.88\% & 99.36\% & 99.30\% \\
		& & f1-score & 81.85\% & 99.39\% & 99.41\% \\
		\hline
        & & recall & 88.60\% & 99.76\% & 99.77\% \\
		& galaxy & precision & 98.29\% & 99.82\% & 99.87\% \\
		avg & & f1-score & 93.19\% & 99.79\% & 99.82\% \\
        \cline{3-6}
		& & recall & 95.54\% & 99.50\% & 99.62\% \\
		& star & precision & 74.33\% & 99.30\% & 99.34\% \\
		& & f1-score & 83.60\% & 99.40\% & 99.48\% \\
		\hline
\end{tabular}
\caption{Difference of evaluation criteria by using multiple modes under different divisions of the training set and test set proportions}
\label{table-1}
\end{table}

\subsection{Blue/Red Band Missing Data}
        
In some situations, the catalog may be incomplete with missing red or blue bands. To test the model's performance on the band loss data, we identified samples with missing red band ($z$, $y$ bands) or missing blue band ($NUV$, $u$ bands) in the test set, respectively.
        
\begin{figure}[t]
\centering
	\includegraphics[width=0.9\linewidth]{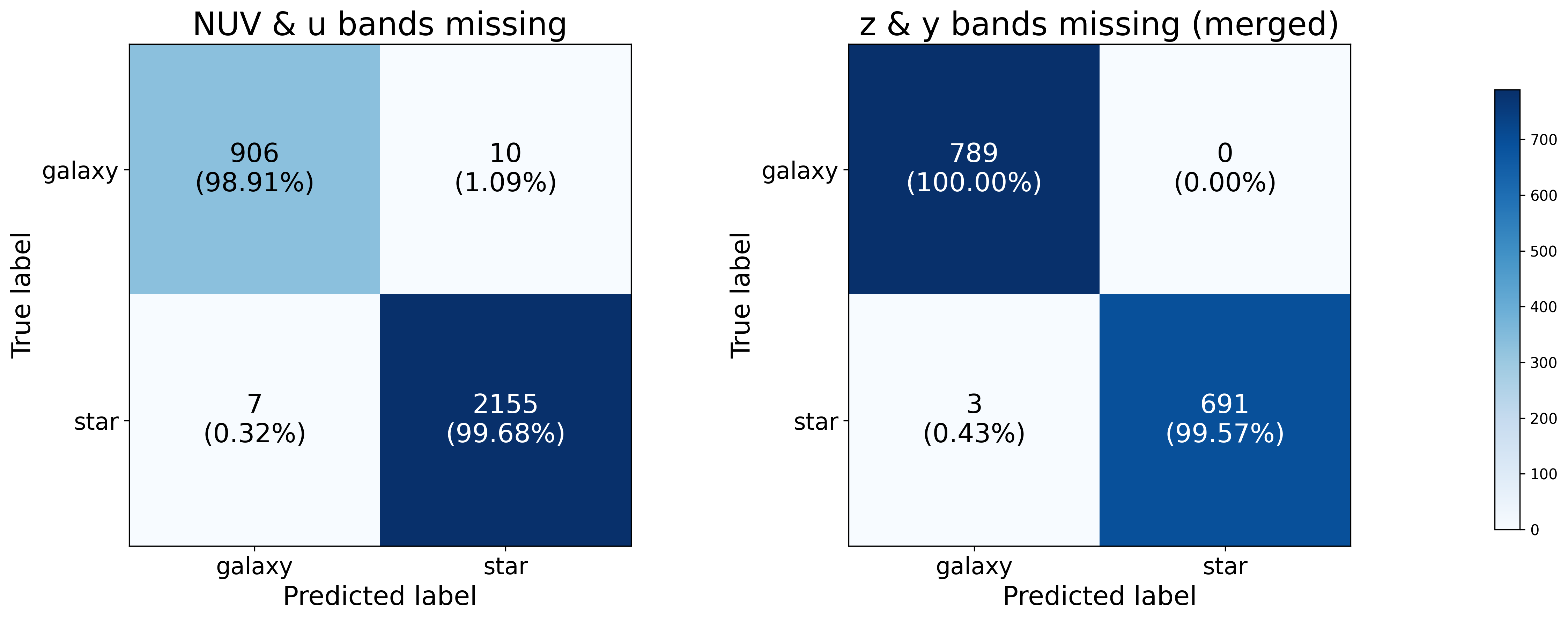}
	\caption{Confusion matrix for the data missing red or blue bands. Left figure: Confusion matrix of the samples that are missing the $NUV$- and $u$-band images and catalogs. Right figure: Confusion matrix of the samples that are missing the $z$- and $y$-band images and catalogs.}\label{fig00-6-3}
\end{figure}

 In Fig. \ref{fig00-6-3}, the left panel is the confusion matrix missing the blue band, and the right panel is the confusion matrix missing the red band. In the samples without data augmentation, there are 9,107 galaxies and 21,559 stars missing the blue band, and 7,942 galaxies and 22 stars missing the red band, respectively. Considering that the sample of stars with missing red band is extremely limited, we selected the samples without missing the blue band and manually removed their images and magnitude of the $z$, $y$ bands. These samples are added to the test set to effectively evaluate classification performance of stars without the red band. The general classification accuracy of the test samples is above 98\%, which demonstrates that our model has great adaptability to missing data samples.
        
\subsection{Performance at Different Magnitudes}
The signal-to-noise ratio (SNR) of the objects can affect the accuracy of classification. For some inherently dim sources or distant galaxies, structures such as bars and disks will be submerged in the noise, resulting in the loss of morphological information and making them easily confused with stars.
        
\begin{figure}[t]
    \centering
    \includegraphics[width=0.7\linewidth]{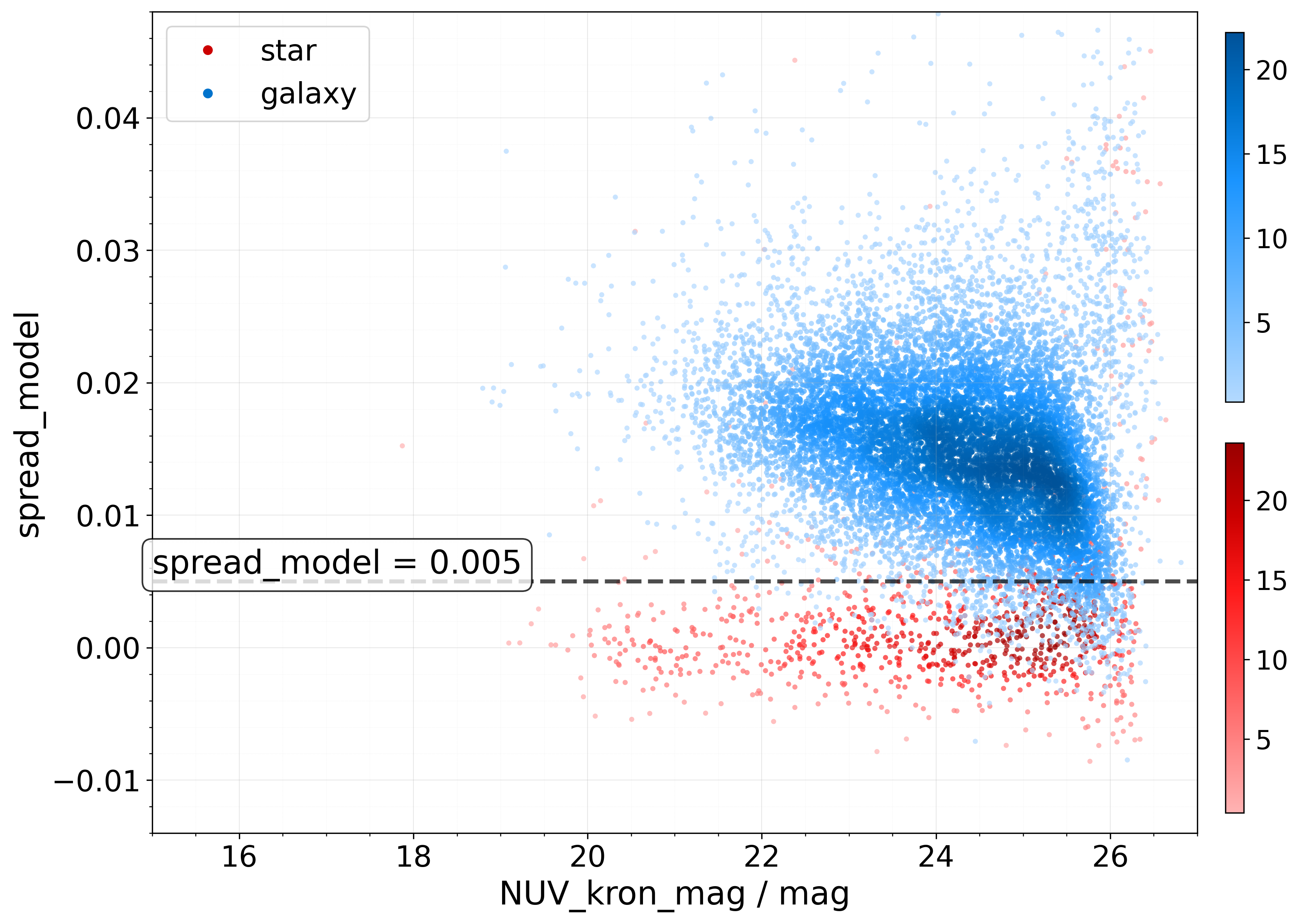}
    \caption{The variation of $spread\_model$ with respect to $kron\_mag$. The horizontal axis represents the $NUV$ band magnitude of a brick, while the vertical axis shows the $spread\_model$ parameter calculated by SExtractor. Star and galaxy samples are represented by red and blue points, respectively. The legend indicates the source density, with darker areas representing a greater number of sources.}
    \label{fig00-7-0}
\end{figure}

\begin{figure}[t]
    \centering
    \includegraphics[width=1.0\linewidth]{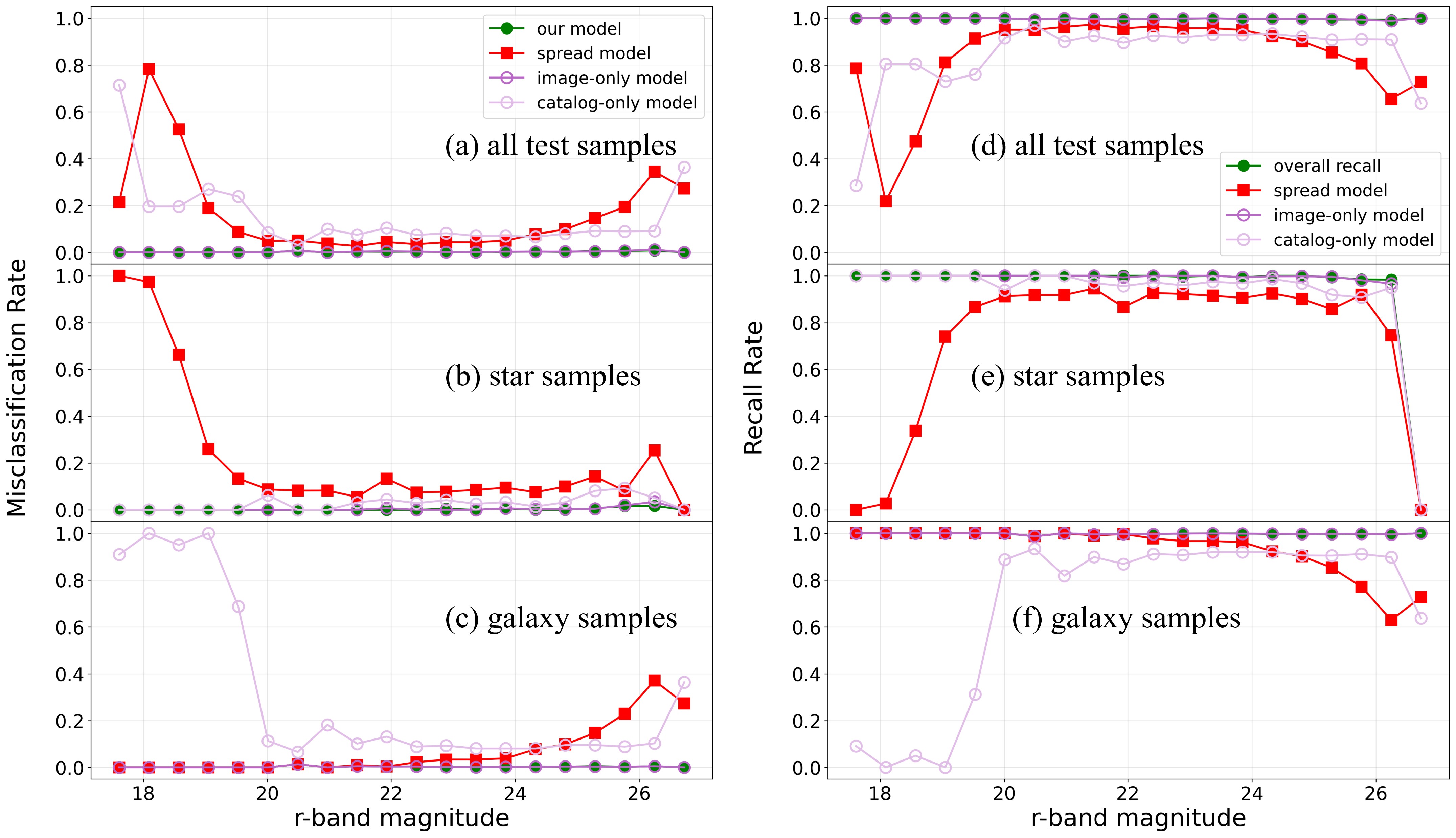}
    \caption{The comparison of the classification effect of the network and spread model based on the variation of r-band magnitude. The red line represents the result of the spread model, while the green line represents ours. And the dark purple and light purple lines respectively represent image-only model and catalog-only model. Panels (a), (b), and (c) respectively show the misclassification rates of all test samples, star samples, and galaxy samples, while figures (d), (e), and (f) show their recall rates.}
    \label{fig00-7}
\end{figure}

We analyzed the error rate of the test samples as the brightness of the stars gradually decreased and compared the results of our model with the spread model. The spread model is a crucial factor in the SExtractor, calculated by comparing the photometric distribution of the object with the matched degree of the PSF model \citep{24} created by the CSST pipeline based on PSFEx \citep{2011ASPC..442..435B}. The spread model could be used as a threshold for star-galaxy separation. While the value is approximately equal to 0, it indicates that the photometric distribution of the object matches the PSF model well, and the object is regarded as a star. Otherwise, it will be regarded as a galaxy. Generally, researchers take 0.005 as the threshold for discrimination \citep{26}. The variation of $spread\_model$ with respect to $kron\_mag$ can be seen in Fig. \ref{fig00-7-0}. We compared the performance of our models with the spread model, the results are shown in Fig. \ref{fig00-7}. Among all test samples, the fusion model exhibits a superior performance to the catalog-only model and outperforms the image-only model in some magnitudes, such as 22 and 26 mag. In the entire brightness range, it also has a lower misclassification rate than the spread model method, especially for brighter and weaker objects. 
        
As we mentioned before, the galaxy will appear more compact while the magnitude is dimmer, which makes it more difficult to distinguish it from the star. The error rate will also gradually increase as the sources become dimmer. When the sources are weaker than 23 mag, our model can still maintain an extremely low misclassification rate, while the error rate of the spread model method gradually increases and reaches 30\% in 26 mag. It can also be seen from Fig. \ref{fig00-7-0} that when the magnitude is weaker than 22, stars and galaxies become indistinguishable. As for the case during 17 and 19 mag, the spread model method has almost no effect, and the misclassification rate even reaches an horrendous 80\%. This might be due to the saturation overflow of the detector in receiving photons, which masks the significant luminosity gradient at the center, causing the object to resemble a point source, which led to the confusion of stars and galaxies. 

\subsection{Performance at Different Redshifts}

While the redshift of galaxies increases, their surface brightness decreases according to the function $(1+z)^{-4}$ according to the Tolman effect \citep{27}. This results in higher redshift galaxies appearing darker and more blurry in the image, and their morphological features becoming even more difficult to distinguish from the stars.
        
\begin{figure}[t]
    \centering
    \includegraphics[width=0.7\linewidth]{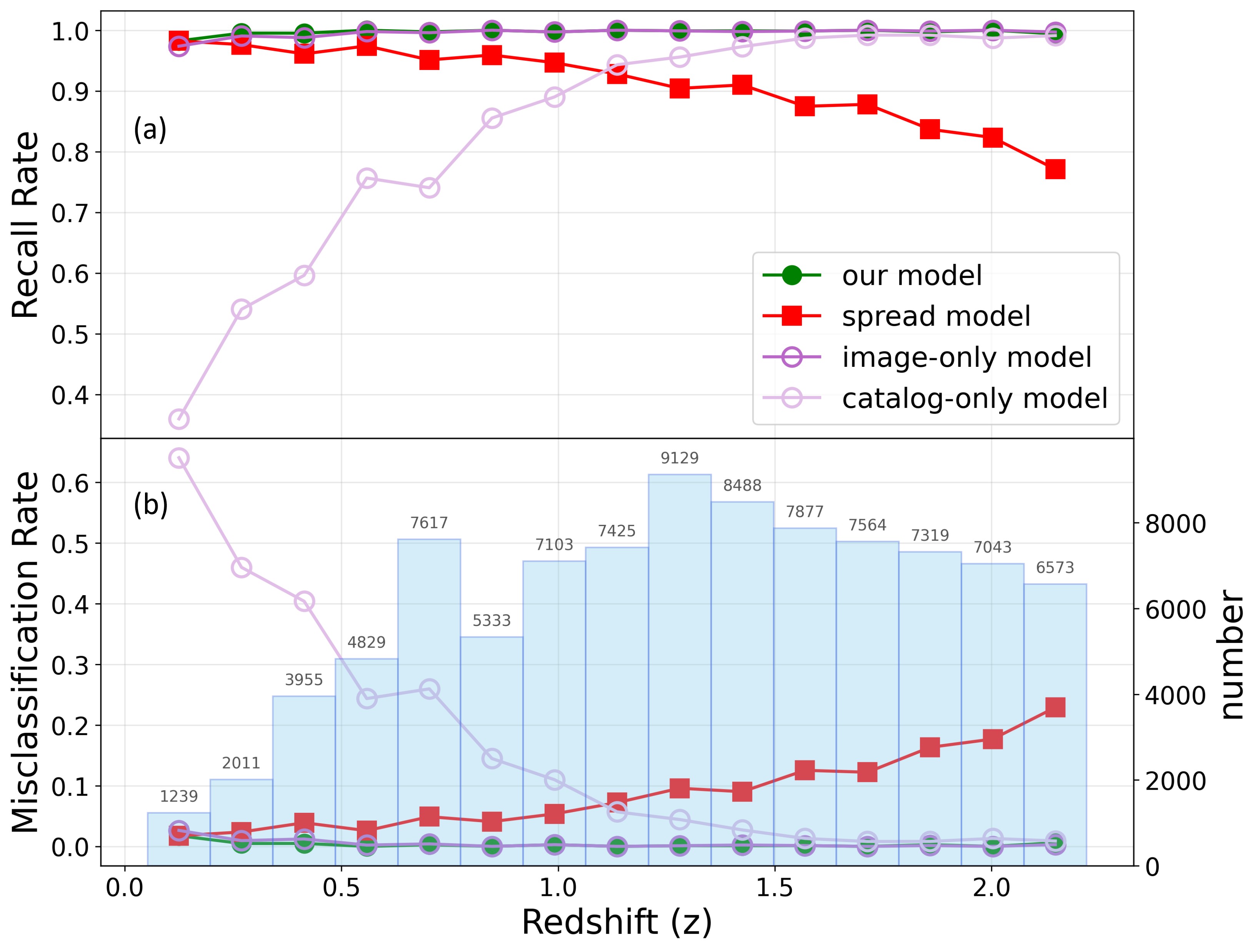}
    \caption{The comparison of the classification effect of the network and spread model based on the increase of redshift. Figures (a) and (b) illustrate the differences in recall rate and misclassification rate, respectively. In Figure (b), the scale on the right side of the vertical axis shows the number of galaxies in the dataset at different redshifts.}
    \label{fig00-7-2}
\end{figure}
        
As shown in Fig. \ref{fig00-7-2}, both the fusion model and the image-only model outperform the spread model, and the fusion model gets the best performance due to its lower misclassification rate in z < 0.5. The misclassification rate of the spread model gradually increases from around 0.1 to 2.2 as the redshift increases, eventually exceeding 20\%, while the fusion model maintains a very low misclassification rate around 0.5\%. This result can be attributed to the network's robust feature extraction capacity and the simulation dataset's comprehensive high-redshift galaxy coverage, which together maintain robustness across the redshift range of 0.1 to 2.2.

The catalog-only model performed poorly overall, and the misclassification rate decreased as the redshift increased. This is because only the magnitude of 7 bands are used for classification, which cannot provide effective feature information, especially in redshift bins with relatively small sample sizes. This issue is also reflected in Figure \ref{fig00-7}, where the high misclassification rate of the galaxy samples is evident within the 18 to 20 magnitude range.

\section{Conclusion}

In this paper, we propose a multimodal data fusion model based on ResNet-50 and BiLSTM networks to classify stars and galaxies in CSST simulation data. The principal conclusions derived from the experiments are summarized below.

    1. Our model achieves high recall and precision in classifying stars and galaxies. The galaxies achieved a recall rate of 99.81\% and a precision rate of 99.88\%, while the stars achieved 99.66\%, 99.44\%, respectively. The general classification accuracy was above 99.75\%. This shows that our model has a strong ability to distinguish between stars and galaxies, which is helpful for later celestial classification of the CSST measured data to provide pure samples for scientific research.

    2. The model maintains high performance for objects with missing blue or red band samples. For galaxies, the recall rate remains above 98\% when missing the blue or red band. For stars, the recall rate for stars reaches a higher 99.5\%.

    3. Experimental results demonstrate that our model exhibits strong robustness, as evidenced by its stable misclassification and recall rates independent of magnitude, coupled with its maintained fine classification performance for high-redshift galaxies, with a misclassification rate below 0.5\% for redshifts between 1 and 2. In contrast to our model, the traditional method based on the spread model from SExtractor shows a high rate of confusion ($\approx$30\%) for stars and galaxies that are dimmer than 23 mag, and the misclassification rate increases significantly with redshift, reaching 20\% at z=2. These results demonstrate that our model is particularly well-suited for classifying dim sources and high-redshift galaxies. This capability will facilitate effective classification of data from upcoming CSST observations.
	
	These results prove that the RBiM network has a strong effect on the classification of stars and galaxies, especially faint sources and high-redshift galaxies. After some adjustments, it will be an applicable model for the classification of CSST data obtained in the future. Despite the greater diversity and complexity of galaxy morphologies expected in CSST measured data, our model could be effectively adapted with only a minimal amount of such data for fine-tuning specific layers via transfer learning. Compared to simulation images, measured images will contain more noise and interference sources. The image pre-processing stage also represents an area for future enhancement, which will be addressed in subsequent work.

\section{Acknowledgements}
This work is based on the mock data created by the CSST Simulation Team, which is supported by the CSST scientific data processing and analysis system of the China Manned Space Project. The simulated data were reduced by the CSST scientific data processing and analysis system of the China Manned Space Project. This work is supported by China Manned Spaced Project (CMS-CSST-2025-A21) and the National Natural Science Foundation of China (NSFC; grants 12233008).

\appendix
\section{Model Training Visualizations}

\subsection{t-SNE Visualizations of Feature Space}
\begin{figure}[t]
    \centering
    \includegraphics[width=0.7\linewidth]{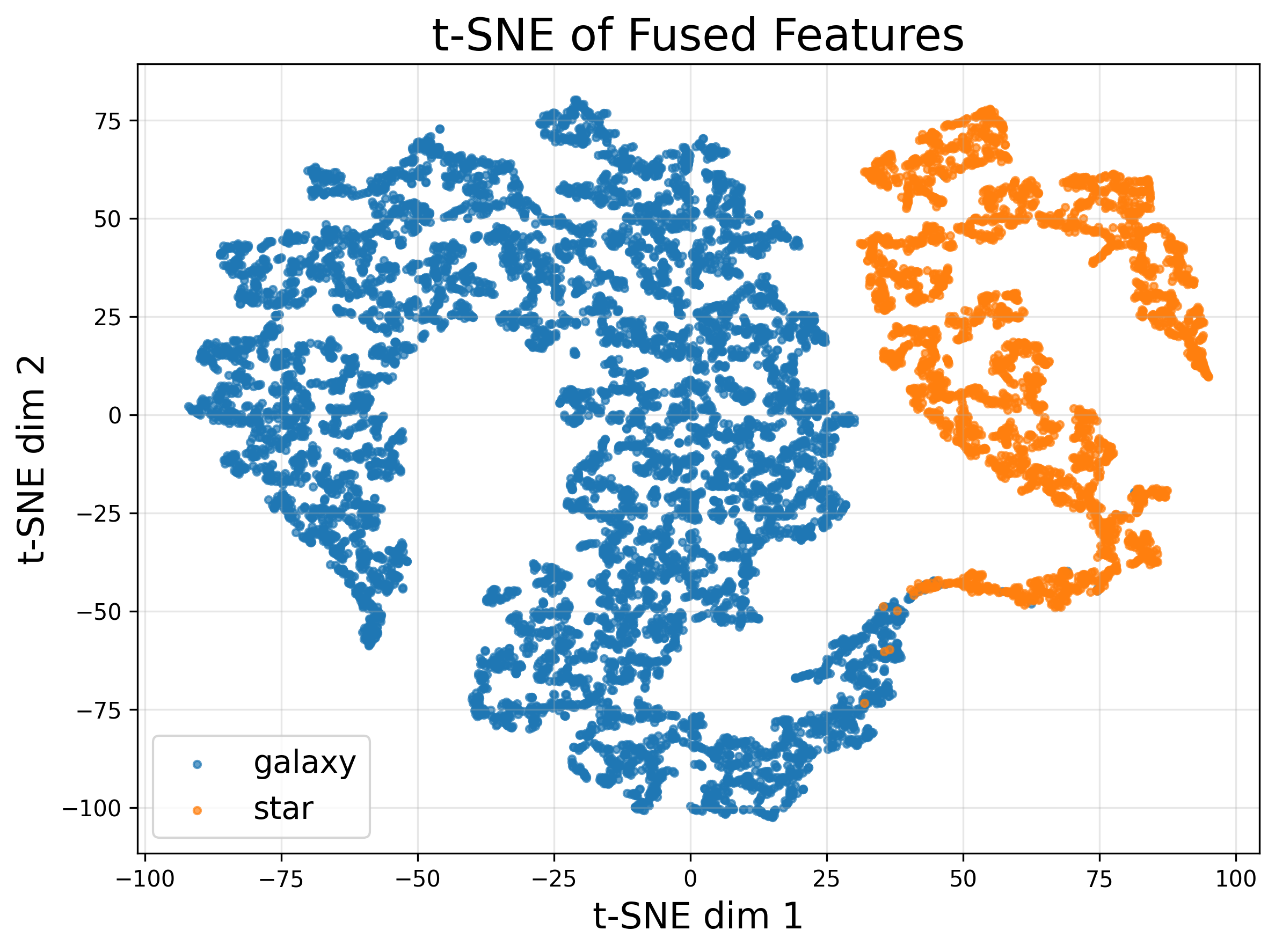}
    \caption{t-SNE two-dimensional feature map. The blue points are galaxy samples, while the orange points are star samples.}\label{fig:t-SNE}
\end{figure}

The non-linear dimensionality reduction algorithm t-SNE is employed to reduce the fused high-dimensional features to a two-dimensional space. As shown in Figure \ref{fig:t-SNE}, the star and galaxy samples were well separated, with only a few confused samples. This indicates that the model has learned effective features to distinguish stars from galaxies. 

\subsection{ROC Curves and Performance Evaluation}
\begin{figure}[t]
    \centering
    \includegraphics[width=0.7\linewidth]{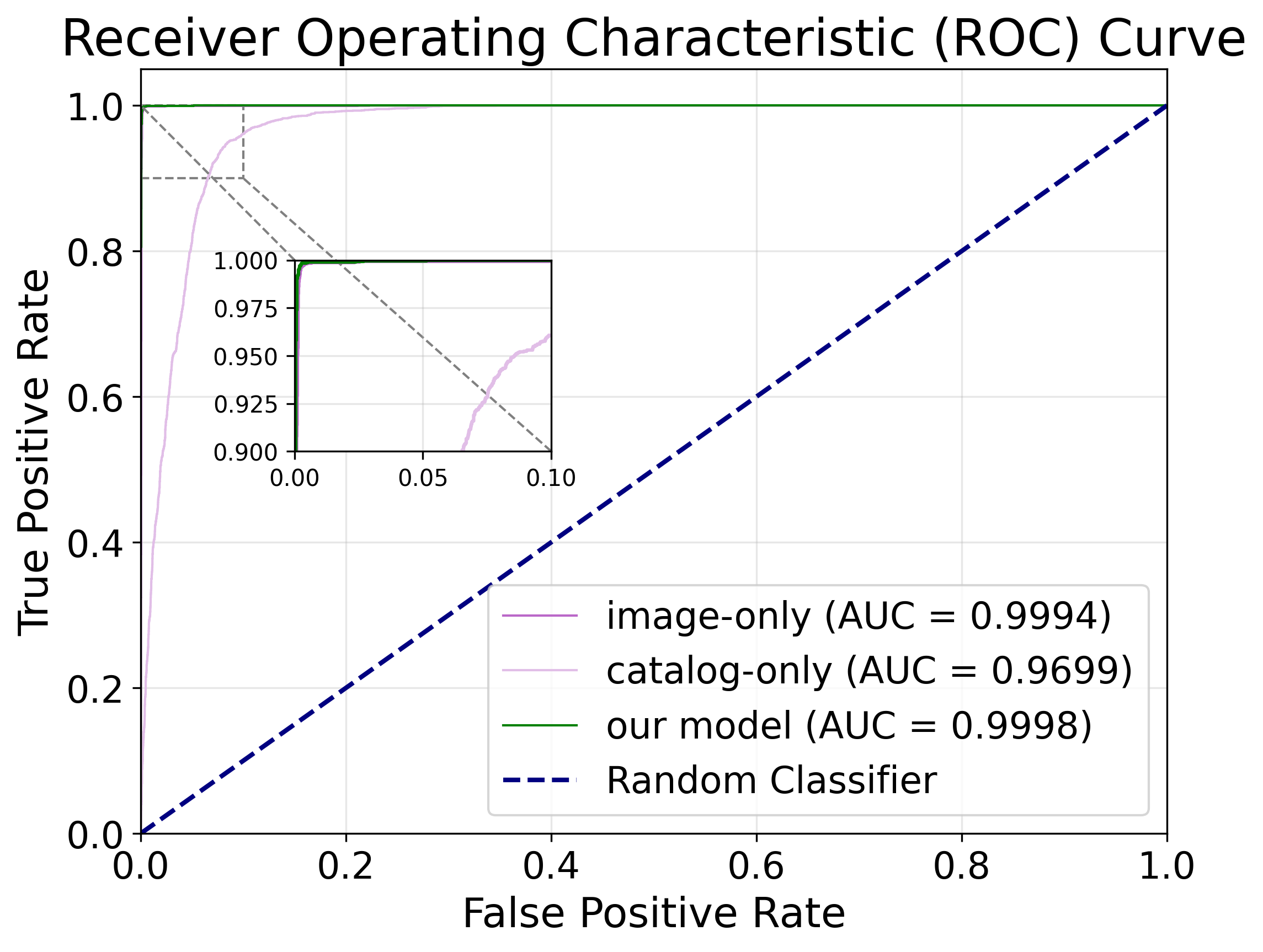}
    \caption{ROC curves. In the enlarged subgraph, the fusion model is sharper at the corners and achieves a larger AUC area.}\label{fig:ROC}
\end{figure}

To further verify the interpretability and effectiveness of the model, we plotted the Receiver Operating Characteristic (ROC) curve. Figure \ref{fig:ROC} presents the ROC curves and their corresponding Area Under Curve (AUC) for the catalog-only, image-only, and fusion model. We can observe that these three models achieve good performance in the classification tasks of stars and galaxies (AUC > 0.95). The fusion model achieves the highest classification accuracy among the three models. It demonstrates a clear superiority in both ROC and AUC over the catalog-only model, while also modestly exceeding the performance of the image-only model.

\label{appendix-Figures}

\end{document}